\documentclass[aps,pra,superscriptaddress,reprint]{revtex4-1}
\usepackage{amsmath}    
\usepackage{graphicx}   
\usepackage[dvipdfx,cmyk]{xcolor}     
\usepackage[colorlinks,linkcolor=red,citecolor=brown,urlcolor=blue]{hyperref}
\usepackage{dsfont,bm}
\usepackage{color}
\usepackage{placeins}
\usepackage[T1]{fontenc} 
\definecolor{dgreen}{RGB}{0,120,0}

\usepackage{amsmath} 
\usepackage{amsthm} 
\usepackage{amssymb}	
\usepackage{graphicx} 
\usepackage{mathtools} 

\newcommand{\ket}[1]{\left| #1 \right>} 
\newcommand{\bra}[1]{\left< #1 \right|} 
\let\baraccent=\= 
\renewcommand{\=}[1]{\stackrel{#1}{=}} 

\theoremstyle{definition}

\theoremstyle{remark}

\DeclarePairedDelimiter{\ceil}{\lceil}{\rceil} 

\graphicspath{{Figures/}}

\begin{document}
\title{Multiplexed entanglement generation over quantum networks using multi-qubit nodes}
\author{Suzanne B. van Dam}
\thanks{These authors contributed equally}
\author{Peter C. Humphreys}
\thanks{These authors contributed equally}
\affiliation{QuTech, Delft University of Technology, PO Box 5046, 2600 GA Delft, The Netherlands}
\affiliation{Kavli Institute of Nanoscience, Delft University of Technology, PO Box 5046, 2600 GA Delft, The Netherlands}
\author{Filip Rozp\k{e}dek}
\thanks{These authors contributed equally}
\affiliation{QuTech, Delft University of Technology, PO Box 5046, 2600 GA Delft, The Netherlands}
\author{Stephanie Wehner}
\affiliation{QuTech, Delft University of Technology, PO Box 5046, 2600 GA Delft, The Netherlands}
\author{Ronald Hanson}
\thanks{r.hanson@tudelft.nl}
\affiliation{QuTech, Delft University of Technology, PO Box 5046, 2600 GA Delft, The Netherlands}
\affiliation{Kavli Institute of Nanoscience, Delft University of Technology, PO Box 5046, 2600 GA Delft, The Netherlands}

\begin{abstract}
Quantum networks distributed over distances greater than a few kilometers will be limited by the time required for information to propagate between nodes. We analyze protocols that are able to circumvent this bottleneck by employing multi-qubit nodes and multiplexing. For each protocol, we investigate the key network parameters that determine its performance. We model achievable entangling rates based on the anticipated near-term performance of nitrogen-vacancy centres and other promising network platforms. This analysis allows us to compare the potential of the proposed multiplexed protocols in different regimes. Moreover, by identifying the gains that may be achieved by improving particular network parameters, our analysis suggests the most promising avenues for research and development of prototype quantum networks. 
\end{abstract}

\maketitle

Recent progress in the generation, manipulation, and storage of distant entangled quantum states has opened up an avenue to the construction of a quantum network over metropolitan-scale distances in the near future~\cite{hensen2015loophole,hucul2015modular}. One of the key challenges in realizing such quantum networks will be to overcome the communications bottleneck induced by the long distances separating nodes. This occurs because probabilistic protocols require two-way communication and, for such distances, the entanglement generation rate becomes limited by the time required for quantum and classical signals to propagate.

It is unlikely that quantum networks will attain sufficient levels of complexity in the near future to support the transmission of complex multi-photon entangled states necessary to overcome this bottleneck through error correction~\cite{munro2015inside, muralidharan2016optimal}. This motivates the development of alternative methods to circumventing this limited communication rate, of which the most promising near-term approach is through multiplexing entanglement generation~\cite{Collins2007,Simon2007,Sangouard2009,Munro2010, sinclair2014spectral,Vinay2016}.

Previous proposals have developed multiplexed entanglement-generation protocols for networks based on atomic-ensemble quantum memories and linear optics~\cite{Simon2007, RevModPhys.83.33, sinclair2014spectral} and for networks in which each node consists of many optically accessible qubits that can be temporally, spectrally or spatially multiplexed~\cite{Collins2007,Sangouard2009,Munro2010,Vinay2016}. However, these proposals are not effective for promising multi-qubit hybrid network node architectures~\cite{nickerson2013topological}, in which one (or a few) optically accessible communication qubits in each node provide a communication bus to interface with multiple local memory qubits. Several platforms have demonstrated the key elements of such a system, including nitrogen-vacancy (NV) centres in diamond \cite{Bernien2013,Gao2015}, trapped ions~\cite{hucul2015modular}, and quantum dots\cite{Gao2015,Delteil2016}.

\begin{figure}[htbp]
\begin{center}
\includegraphics[width=8.0cm]{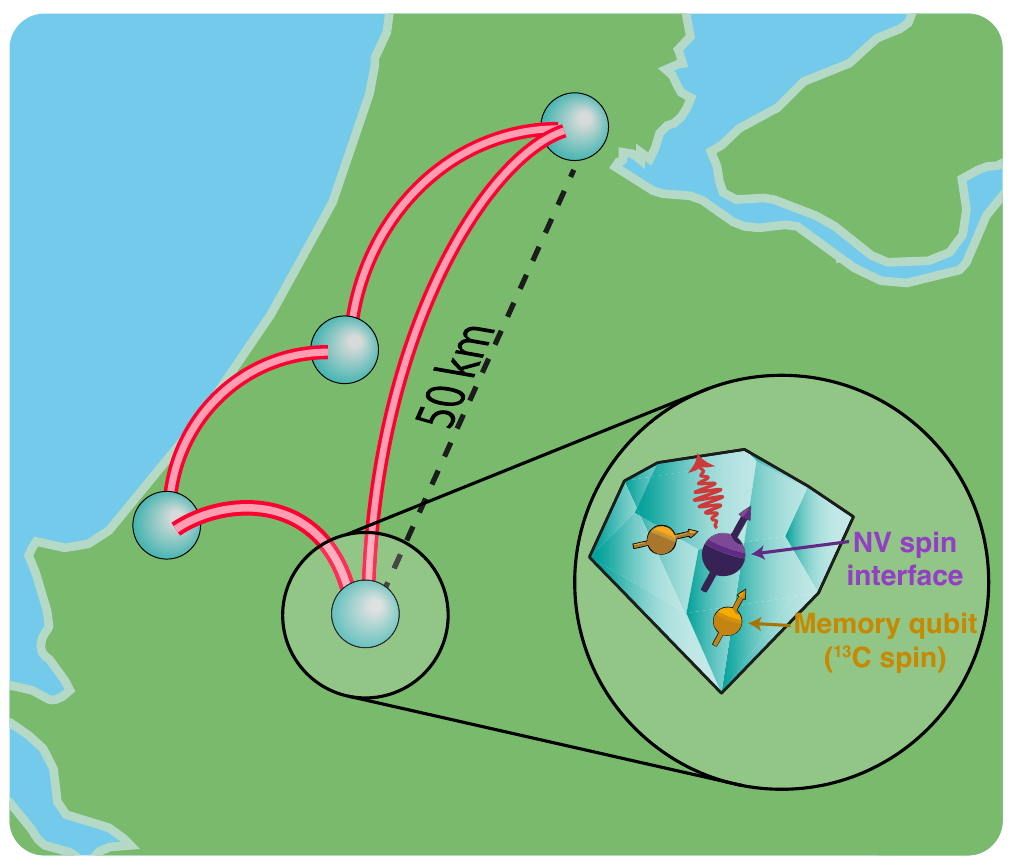}
\caption{Quantum networks have the potential to reach metropolitan scales in the near term, opening up new challenges due to the time required to signal successful entanglement generation between nodes separated by many kilometers. Nitrogen-vacancy centres in diamond are promising candidates for the nodes of such a network, combining an electronic spin communication qubit interface for entanglement generation and local processing with long lived $^{13}$C nuclear-spin memory qubits. }
\label{fig:Network}
\end{center}
\end{figure}

Here we focus on the scenario of efficiently generating heralded remote entanglement between two hybrid multi-qubit nodes separated by tens of kilometers in a quantum network (Fig.~\ref{fig:Network}). We propose two strategies for multiplexing entanglement generation using multi-qubit architectures, identifying the scaling of the entangling rates with the distance between nodes. We compare these strategies to an alternative protocol based on the distribution of entangled photon-pairs \cite{Jones2016}, modelling all three protocols analytically and with Monte Carlo simulations. This allows us to identify optimal protocols for different regimes of distance and node performance. 

In order to be able to effectively assess the potential of these network protocols, it is vital to incorporate the known and anticipated limitations of potential platforms from the start. In this paper we therefore use network parameters representing the expected near-term performance of NV centre nodes. These centres are promising nodes for such a network, combining a robust and long-lived $^{13}$C nuclear-spin quantum register~\cite{Cramer2016,PhysRevX.6.021040} with a photonic interface (Fig.~\ref{fig:Network}). Our conclusions are nonetheless broadly applicable to other platforms with comparable system performances, particularly including trapped ions~\cite{hucul2015modular}.

\section{Quantum network protocols}\label{sec:protocols} 
\label{sec:protocols}
We begin by briefly introducing the three candidate protocols that we consider for a metropolitan-scale quantum network. For each network, we identify the scaling of the entanglement generation rate with the system transmission efficiency and the distance between nodes.

\subsection{Multiplexed Barrett-Kok protocol}

The first scheme is a multiplexed version of the Barrett-Kok (BK) protocol. In this scheme, entanglement is generated at both nodes locally between the spin state of the communication qubit and the modal occupation of a single photon (typically the photonic state is time-bin encoded for NVs). This procedure constitutes a single attempt to generate remote entanglement. The two photons are then transmitted to a remote beam splitter, where a probabilistic joint Bell state measurement (BSM) on the photons projects the two distant communication qubits into an entangled state upon measurement of the appropriate outcomes~\cite{PhysRevA.71.060310}.

In this protocol each photon needs to be transmitted over a distance $d/2$ from the nodes to the central BSM station. This is followed by the transmission of classical information over the same $d/2$ distance heralding to the nodes the success or failure of the entangling attempt. Hence in the standard BK protocol, the entanglement attempt rate $r_{BK}$ is limited by the combined quantum and classical communication time ($t_c = d/c$) required to establish whether the protocol succeeded: $r_{BK} \sim t_c^{-1}$. Even for modest distances, this time delay is sizable; e.g. for $d=50$ km the delay is $t_c = 250 \, \mu$s, limiting the attempt rate to 4~kHz. 

\begin{figure}[htbp]
\begin{center}
\includegraphics[width=8.5cm]{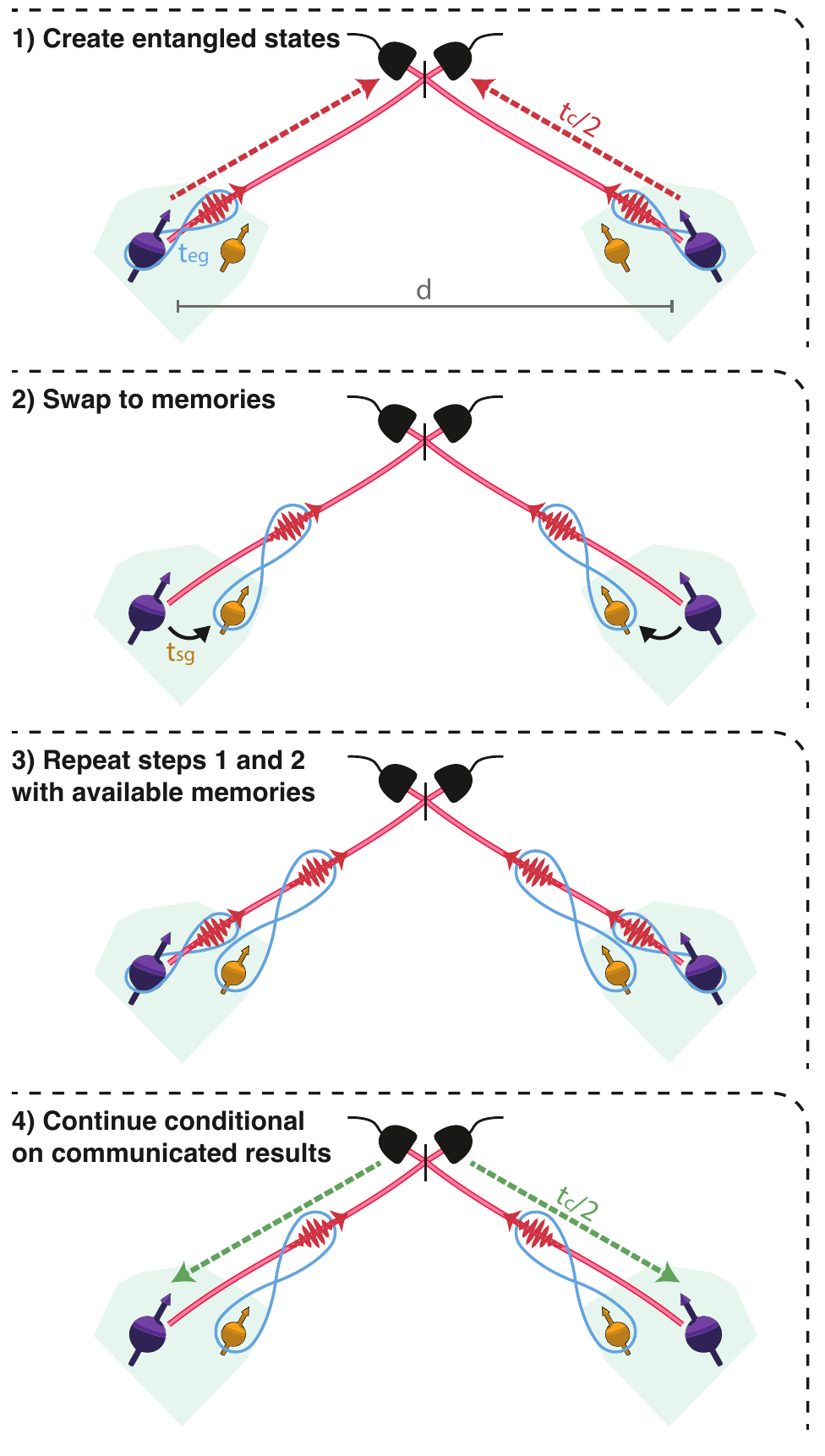}
\caption{Multiplexing concept. The protocol starts with a creation of local entanglement between the communication qubits and single photons at both nodes (step 1). The state of the communication qubits is then immediately transferred to the memory qubits (step 2), which allows for a second entanglement attempt before the result of the first one is known (step 3). Once the signal heralding success or failure of the attempt is received at the nodes, the occupied memories can be reused for new attempts (step 4).}
\label{fig:MultiplexingDiagram}
\end{center}
\end{figure}

This rate limitation can be mitigated by using a multiplexed version of the BK protocol (Fig.~\ref{fig:MultiplexingDiagram}), in which the spin state of the communication qubit is swapped to a memory qubit directly after spin-photon entanglement generation, freeing up the communication qubit for additional entanglement generation attempts. For the NV system, naturally occuring nearby $^{13}$C nuclear spins provide robust memory qubits~\cite{PhysRevX.6.021040, blok2015towards}. The state is stored in this memory qubit until information about the success of the attempt arrives. In the meantime, spin-photon entanglement generation and subsequent state swapping to other memories can continue until all of the memories are occupied. The multiplexed protocol allows $N$ qubits per node to be utilised, where $N$ includes both the communication qubit and the memory qubits.

The maximum number of qubits per node that can be usefully employed in this protocol is given by $N_{\mathrm{max}} = \ceil{t_c/t_{sg}}$~\footnote{$\ceil{x}$ denotes ceil($x$)}  where $t_{sg}$ is the duration of the swap gate (typically much longer than the duration of entanglement generation attempts $t_{eg}$). The attempt rate of the multiplexed Barrett-Kok (mBK) protocol is therefore a factor $N$ larger than for the standard BK scheme: $r_{\text{mBK}} \sim N / t_c$  for $N \leq N_{\mathrm{max}}$. This rate is upper bounded by $r_{\text{mBK}} \leq 1/t_{sg}$.

The success of each attempt of the BK scheme is conditioned on the detection of both the photons emitted by the communication qubits in the BSM. As a result, the system transmission efficiency $\eta$ appears quadratically in the entanglement success rate $R_{\text{mBK}}$. Hence for $N \leq N_{\mathrm{max}}$:
\begin{equation}
R_{\text{mBK}} \sim r_{\text{mBK}} \, \frac{\eta^2}{2} = \frac{1}{2}N \, \eta^2 / t_c.
\end{equation}
The factor of half corresponds to the probability of a successful BSM at the beam splitter.

\subsection{Multiplexed Extreme-Photon-Loss Protocol}
In the case of high levels of photon loss $(\eta \ll 1)$, a protocol based on entanglement distillation can be more effective than the BK protocol. In this protocol, instead of directly trying to generate a maximally entangled state $\ket{\Psi} = (1/\sqrt{2})(\ket{01} + \ket{10})$, two weakly entangled states of the form $\rho \approx \frac{1}{2} \ket{\Psi}\!\bra{\Psi}+ \frac{1}{2} \ket{00}\!\bra{00}$ are efficiently generated conditional on the detection of only a single photon at the beam splitter station~\cite{PhysRevLett.101.130502, nickerson2013topological}. Here $\ket{0}$ ($\ket{1}$) denotes the state of the communication qubit from which a photon is (is not) emitted. These weakly entangled states contain a contribution $\ket{00}\!\bra{00}$ from the case in which both communication qubits emitted a photon, but only one was detected.  After the two states are successfully generated, an entanglement distillation procedure is performed using local operations and classical communication. This distillation produces a pure entangled state with a $1/8$ probability. Since two raw states are consumed to generate a final entangled state, this extreme-photon-loss (EPL) protocol requires at least two qubits per node, as the first state has to be stored in a memory qubit until the second entangled state is generated. 

The advantage of this scheme over the BK protocol is that it does not require the detection of coincident photons, instead allowing for multiple attempts to generate the second state. This results in a success probability that is proportional to $\eta$ rather than $\eta^2$ and thus an entangling rate $R_\text{EPL} ~\sim \eta / (16 t_c)$, where a factor $1/8$ corresponds to the probability that the distillation operation succeeds, and a factor $1/2$ reflects the need to generate two entangled states.

Analogously to the BK protocol, a multiplexed version of the scheme can be envisioned in which multiple entanglement generation attempts are performed within one communication cycle. Since, in the second stage of the protocol one memory is continuously occupied by the first entangled state, the maximum number of qubits that can be effectively utilised is one more than in the BK protocol:  $N_{\mathrm{max}} = \ceil{t_c/t_{sg}} + 1$. The resulting entanglement success rate $R_\text{mEPL}$ for the multiplexed extreme-photon-loss protocol for $N \leq N_{\mathrm{max}}$ is proportional to the inverse of the sum of the time spent in the first stage $(t_c/(\eta N))$ and second stage $(t_c/(\eta (N-1)))$ of the protocol: 
\begin{equation}
R_\text{mEPL} \sim \, \frac{N(N-1)}{2N-1} \frac{\eta}{8 t_c}.
\label{eq:rateEPL}
\end{equation}

The entangled state fidelity in this protocol is sensitive to decoherence of the memories during entanglement attempts. In order to ensure a minimum fidelity, stored entangled states can be discarded after a set number of subsequent entanglement attempts, at the expense of decreasing the entanglement rate. Entanglement generated from a single photon detection event is expected to succeed within at most a few hundred attempts ($\sim$100 attempts at 50 km, $\sim$1000 attempts at 100 km) for the range of parameters considered here. For nitrogen-vacancy centre nodes, recent results indicate that $^{13}$C nuclear-spin memories may effectively preserve quantum states over this number of attempts~\cite{PhysRevX.6.021040}, and so this effect is not expected to significantly impact our conclusions.

\subsection{Midpoint-Source Protocol}

\begin{figure}[htbp]
\begin{center}
\includegraphics[width=8.5cm]{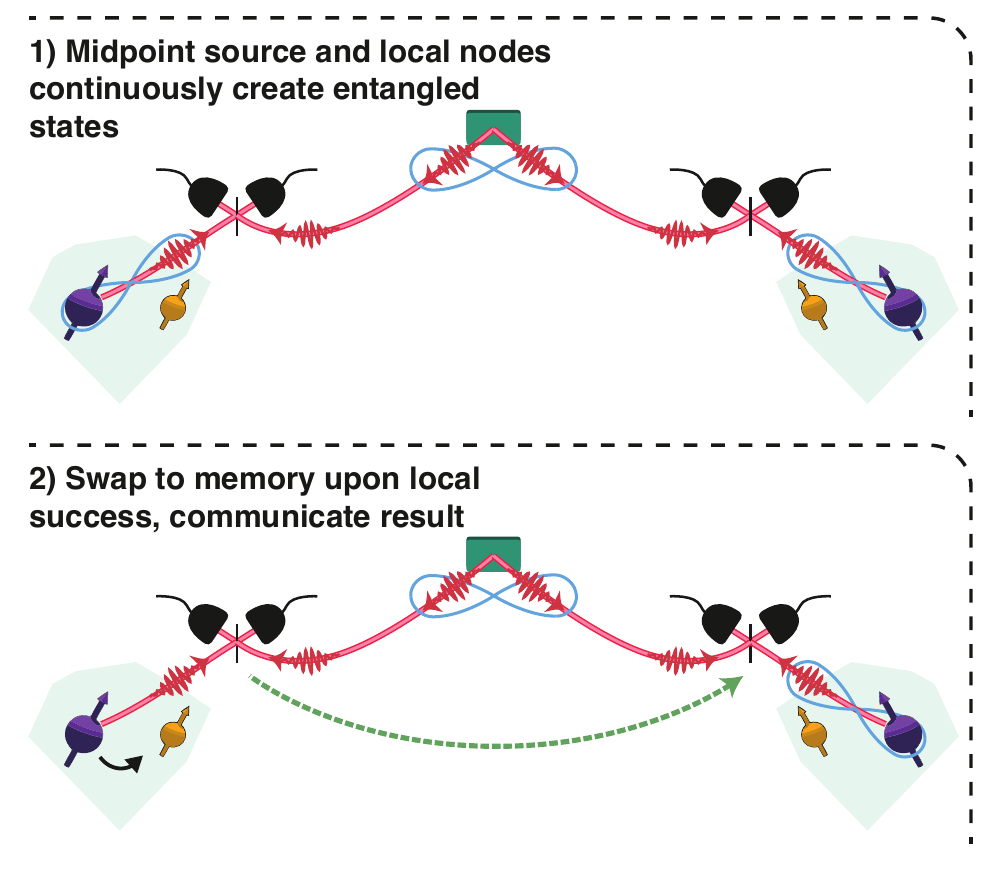}
\caption{Midpoint-source (MPS) protocol. The photon source in the middle continuously generates pairs of entangled photons with probability $p_\text{em}$ and transmits them to the two nodes (step 1). At the same time both nodes synchronously generate local entanglement between the communication qubit and emitted photons. Local beam splitter stations at each node perform BSM measurements between photons emitted from the source and the the photons emitted from the local node. This gives the local node immediate knowledge of the local success or failure of each attempt. This information is also communicated to the other node, arriving $d/c$ later.}
\label{fig:MidPointSource}
\end{center}
\end{figure}

The final configuration that we consider is the midpoint-source (MPS) protocol following Ref. ~\cite{Jones2016}. In addition to the two nodes, this protocol requires an entangled-photon source (which emits pairs of photons with probability $p_\text{em}$) positioned midway between the nodes (Fig. \ref{fig:MidPointSource}). In this protocol, pairs of entangled photons generated by the photon source are split and one is sent to each of the two nodes. At each of the nodes, a BSM is performed between this photon and a photon generated by the local communication qubit. Entanglement swapping succeeds only if both BSMs succeed (requiring the detection of four photons in total). 

Since the successes of the BSMs can be reported to their local nodes immediately, in the case of local failure the nodes can quickly proceed to a new entanglement generation attempt. In this way the entanglement attempt rate can be significantly increased. The attempt rate is upper bounded by $r_{\text{MPS}}\leq t_{eg}^{-1}$, where $t_{eg}$ is the duration of the spin-photon entanglement generation. 

This upper bound is saturated if the number of successful local BSMs per communication time $t_c$, $n = p_{\text{BSM}} \, t_c/t_{eg}  \approx (1/2) p_{em} \eta \, t_c/t_{eg}$, satisfies $n \ll 1$. In this limit the protocol can be effectively run with a single qubit per node, and the rate is therefore insensitive to the swap gate time $t_\text{sg}$. When operating the MPS protocol in this low $n$ regime, the entanglement success rate is given by
\begin{equation}
R_\text{MPS} \sim p_\text{em} \, \eta^2 / (4 t_{eg}),
\end{equation}
where the factor of $1/4$ arises because both BSMs must succeed in the same round, and $\eta$ includes the system losses for both the photon from the entangled photon source and the locally generated photon. 

This scaling is different to that identified in Ref.~\citenum{Jones2016} since, for the system parameters that we consider, $t_\text{eg}$ is not small enough to ensure that the expected number of successes $n$ per communication time $t_c$ approaches unity. As shown in Fig~\ref{fig:Platch_vs_distance}, for a shorter $t_{eg}$, the network could leave this low-success-probability regime. If the attempt rate is high enough to ensure that at least one attempt succeeds locally per $t_c$, the overall entanglement success rate will only primarily depend on whether there was a simultaneous success at the other node; the scaling is thus effectively proportional to $\eta$, which is the scaling described in Ref.~\citenum{Jones2016}. However, achieving this limit clearly requires a shorter $t_\text{eg}$ as the loss (1-$\eta$) increases.

For $n \sim 1$, the inclusion of additional memory qubits becomes beneficial to prevent idle time. In this case, after a local success, the communication qubit state is swapped to a memory qubit. This swapping operation therefore prevents the node from performing further entanglement generation attempts during a time $t_\text{sg}$, limiting the overall attempt rate. 

\section{Modelling}
We model each of the protocols described in the previous section with an approximate analytical approach as well as with Monte Carlo simulations. We use system parameters that are expected to be achievable for NVs and trapped ions in the near term (Tab. \ref{tab:Variables}). The outcoupling efficiency of the NV centre is assumed to benefit from coupling to an optical cavity (with outcoupling efficiency $p_\text{out} = 0.3)$, and emitted photons are assumed to be frequency-converted to telecom-wavelength photons with efficiency $p_\text{fc}=0.3$. Fiber losses are therefore limited to standard telecom values of $\alpha = 0.2$ dB/km. Hence the overall system transmission efficiency is given by $\eta = p_\text{out} \, p_\text{fc} \, 10^{-\alpha d/20}$ where the last term corresponds to the fiber losses over a distance of $d/2$. 

It is as yet unclear how much progress will be made in the near term in overcoming the technical challenges necessary to demonstrate an entangled-photon-source with a high brightness and with spectral properties that are well-matched to the node emission. We therefore consider two possible values for $p_\text{em}$ (0.1 and 0.01), taking 0.01 to be more technically feasible~\cite{clausen2014source,Loredo:16}.

\begingroup
\begin{table}[ht]
\caption{Anticipated near-term parameters for a quantum network based on NV centers \cite{Bernien2013,Cramer2016,Bogdanovic2016,Zaske2012}. These parameters are also anticipated to be achievable using trapped ions~\cite{hucul2015modular}.} 

\begin{center}
\begin{tabular}{|c|p{5.5cm}|c|}
\hline
Variable & Description &  Value \\ \hline
$N$ & Total number of qubits at each node &  2 \\
$p_\text{fc}$ & Frequency-conversion efficiency & 0.3 \\
$p_\text{out}$ & NV-outcoupling efficiency & 0.3  \\
$t_\text{eg}$ & Spin-photon entanglement generation time &1 $\mu$s \\
$t_\text{sg}$ & NV-carbon swap gate time  &  200 $\mu$s \\
$p_\text{em}$ & Midpoint-source photon-pair emission probability  & 0.01, 0.1 \\ \hline
\end{tabular}
\end{center}
\label{tab:Variables}
\end{table}%
\endgroup

\subsection{Scaling with distance}
\begin{figure}[htbp]
\begin{center}
\includegraphics[width=8.5cm]{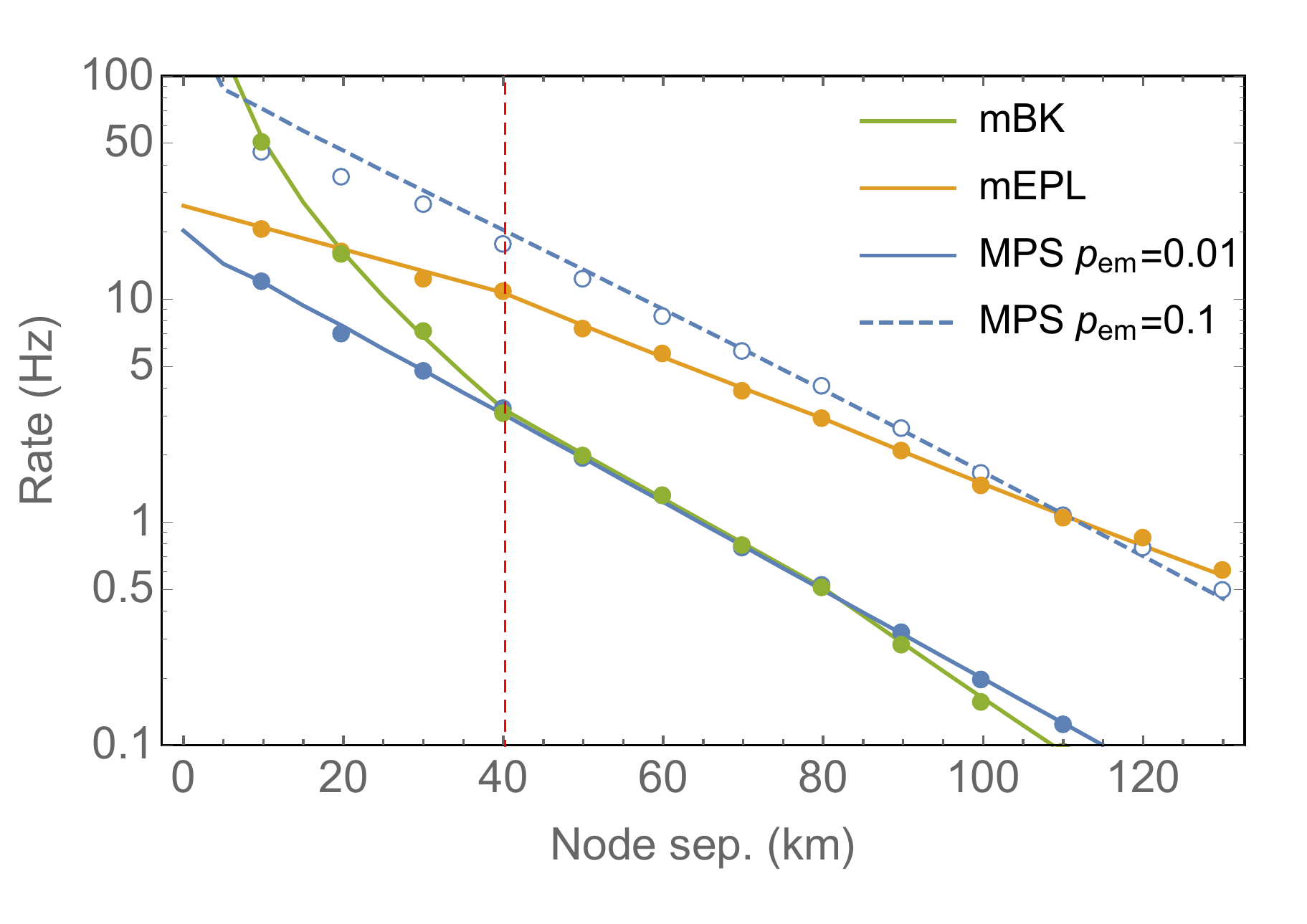}
\caption{Modelled entanglement generation rates as a function of distance for the system parameters listed in Table~\ref{tab:Variables}. Plotted lines give the results of our analytical model while the circles give equivalent Monte Carlo simulation data. Although two qubits are available to the system, the MPS protocol is always found to be in the low success probability regime ($n < 1$), in which only one qubit is required. For distances to the left of the red vertical dashed line the memory storage time $t_\text{sg}$ is larger than the communication time $t_c$. In this regime it is optimal to use only one qubit for the mBK scheme. As the mEPL-scheme requires one memory qubit to store the first generated state in the second part of the protocol, for all distances both qubits are actively employed. The error bars associated with the Monte Carlo simulations are smaller than the plotted circles.}
\label{fig:Rate_vs_distance}
\end{center}
\end{figure}

The modelled dependency of the entangling rate on the node separation is shown in Fig.~\ref{fig:Rate_vs_distance}. As expected from Section~\ref{sec:protocols}, the scaling with distance is most favorable for the mEPL protocol ($R_\text{mEPL} \sim 10^{-\alpha d/20} \, d^{-1} $), whereas the BK protocol scales worst ($R_{\text{mBK}}  \sim 10^{-\alpha d/10} \, d^{-1} $). Even for an MPS protocol with an extremely efficient source ($p_{em} = 0.1$), the mEPL protocol outperforms it for distances greater than $\sim 100$ km since $R_\text{MPS}$ scales less favourably with distance as $R_\text{MPS} \sim 10^{-\alpha d/10}$.

\begin{figure}[htbp]
\begin{center}
\includegraphics[width=8.5cm]{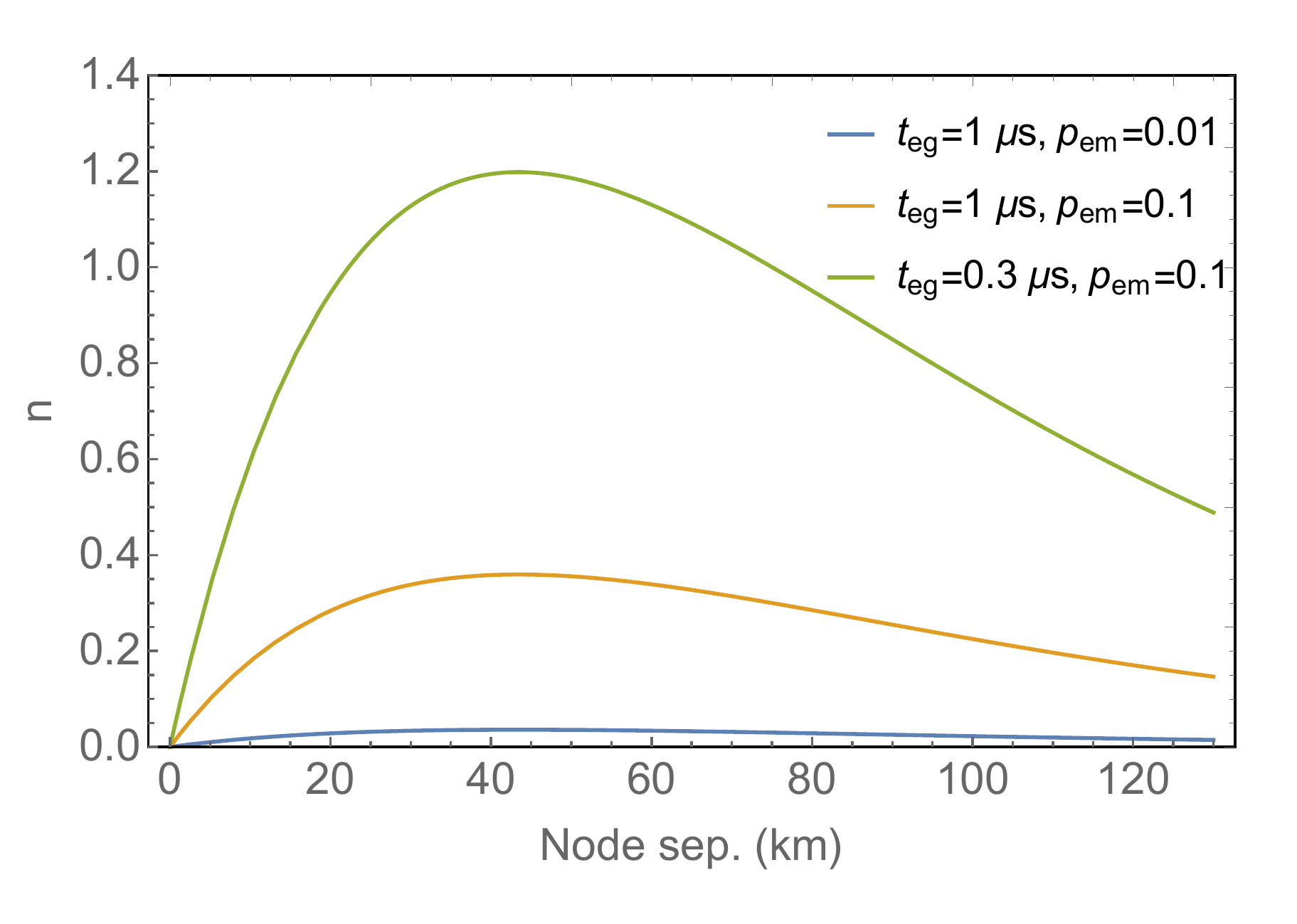}
\caption{Expected number of successful local BSMs $n$ at each node per communication time $t_c$ for the MPS protocol as a function of node separation. We see that for both values of $p_\text{em}$ and for all distances  $n<1$, and hence a single qubit per node is sufficient.}
\label{fig:Platch_vs_distance}
\end{center}
\end{figure}

In Fig.~\ref{fig:Platch_vs_distance} we justify our claim that the MPS protocol will not benefit from more than a single qubit per node. We plot the expected number of successful BSMs $n$ during the communication time as a function of distance, and observe that for our network parameters this stays well below one even for the case of a very efficient source ($p_{em}=0.1$). 

\subsection{Scaling with number of memories}

\begin{figure}[htbp]
\begin{center}
\includegraphics[width=8.5cm]{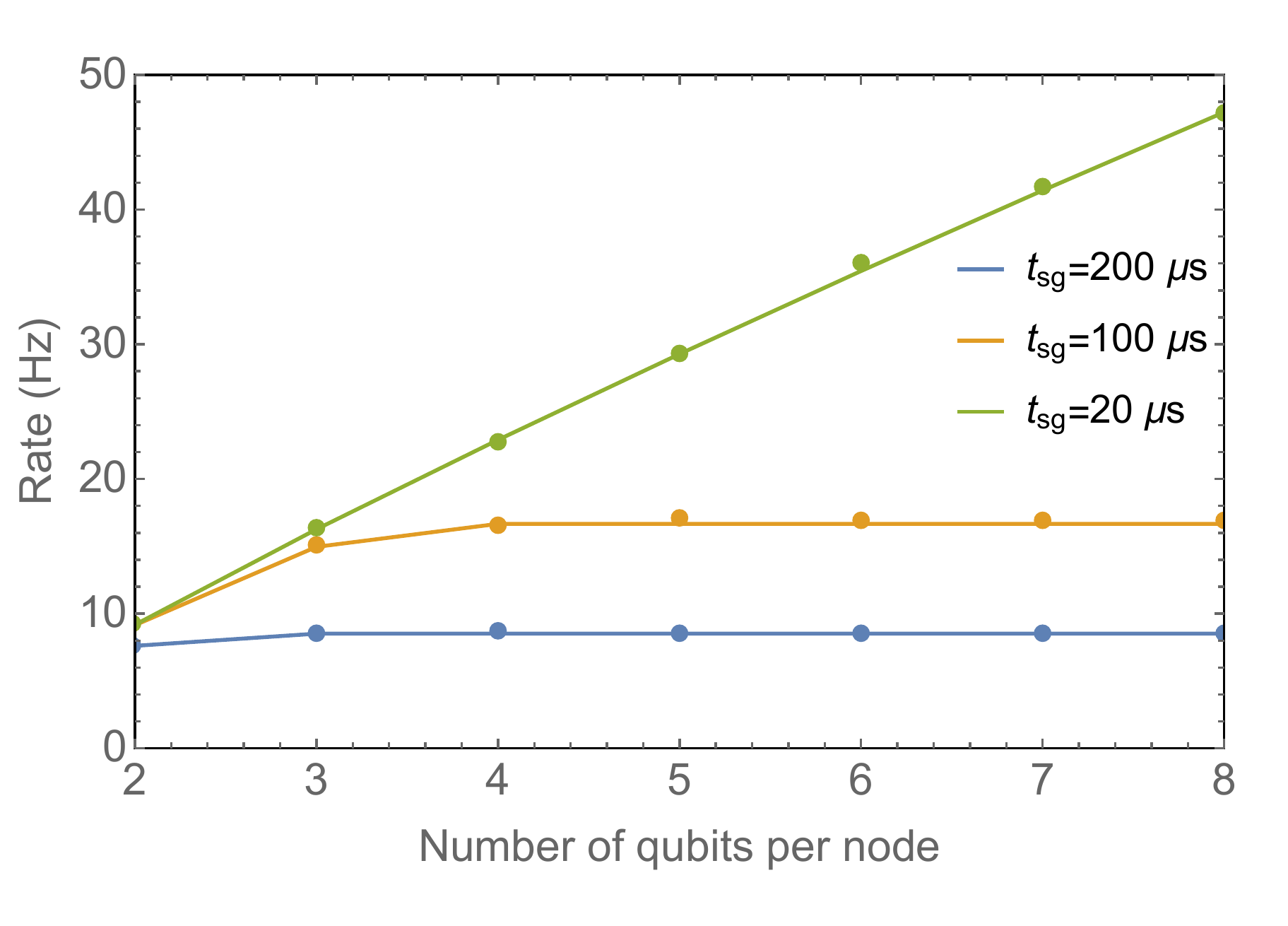}
\caption{Modelled entanglement-generation rate for the mEPL scheme as a function of the number of qubits per node at $d = 50$ km. The three curves correspond to different values of the swap-gate time $t_\text{sg}$. An initial linear scaling of the rate with the total number of qubits is observed, as predicted by Equation~\eqref{eq:rateEPL}. The rate increases only up to $N_{\mathrm{max}} = \ceil{t_c/t_\text{sg}} + 1$, beyond which there is no further benefit. This rate saturation occurs over the addition of two qubits. This is because, while generating the second entangled state in the mEPL protocol, one memory qubit is always occupied by the first generated state. The addition of a further memory qubit beyond $N = \ceil{t_c/t_\text{sg}}$ therefore ensures that there are $\ceil{t_c/t_\text{sg}}$ qubits available for entanglement generation during both phases. However, this memory qubit is only used for the second state generation and so does not contribute as much as previous qubits. Error bars associated with the Monte Carlo simulations are smaller than the plotted circles.}
\label{fig:Rate_vs_NoMem}
\end{center}
\end{figure}

Notably, for these near-term parameters, scaling up to a large number of qubits per node does not speed up the entanglement rate. As previously noted, the MPS protocol always operates in the low success probability regime in which only the communication qubit is actively used. For the mBK and mEPL protocols, the duration of the swap gate significantly limits the number of qubits per node that can be used over relevant node separations. We investigate the rate dependency of the mEPL protocol on the number of memory qubits in Fig.~\ref{fig:Rate_vs_NoMem} for a fixed node separation of $d= 50$ km and a varying swap gate duration $t_\text{sg}$. For $t_\text{sg} \ll t_c$  the rate scales linearly with the number of qubits. However, as explained in Section~\ref{sec:protocols}, once $N t_\text{sg} \approx t_c$ is reached, adding more memory qubits does not boost the entangling rate.

\section{Conclusions}

Our analysis highlights the potential of multiplexed distillation-based schemes to provide high rates of remote entanglement generation and the most favourable scaling with respect to losses. For such schemes, we have identified the swap gate time $t_\text{sg}$ between the communication and the memory qubits as the key parameter in constraining the achievable entanglement generation rate, as this limits the number of quantum memories that can be used. This highlights the importance of developing methods to increase this storage rate while ensuring that memories remain robust to decoherence. One promising approach for nitrogen-vacancy centre nodes may be to use pairs of strongly coupled carbons to encode quantum memories in decoherence protected subspaces that combine rapid gates (due to their strong coupling) with long memory lifetimes~\cite{PhysRevX.6.021040}.

We find that the midpoint-source protocol has a different dependence on the system parameters, with its performance only weakly constrained by the memory storage time. However, its increased sensitivity to losses hinders its performance over long distances. In addition, there is considerable uncertainty in the projected performance of entangled-pair sources in the near-term, particularly with regard to the source brightness. Until brightnesses on the order of 0.1 per attempt can be achieved, our analysis suggests that these schemes will not perform as effectively as the multiplexed distillation-based protocols. 

\begin{acknowledgments}

We acknowledge stimulating discussions with David Elkouss, Kenneth Goodenough, Norbert Kalb, Gl\'{a}ucia Murta, Tim Taminiau and Le P. Thinh. This work was supported by the Dutch Organization for Fundamental Research on Matter (FOM), Dutch Technology Foundation (STW), the Netherlands Organization for Scientic Research (NWO) through a VICI grant (RH), a VIDI grant (SW) and the European Research Council (ERC) through a Starting Grant (RH and SW). 
\end{acknowledgments}

\bibliography{library}
\end{document}